%% file: ZZ_arxiv.tex
\documentclass[twocolumn,prb,superscriptaddress,longbibliography]{revtex4-1}

\usepackage{amsmath,amsfonts,amssymb}
\usepackage{graphicx,color}
\usepackage{hyperref}
\usepackage{epstopdf}
\usepackage{float}
\usepackage{multirow}
\usepackage{hhline}
\usepackage[caption=false]{subfig}

\newcommand{\ket}[1]{\left|#1\right\rangle}
\newcommand{\bra}[1]{\left\langle#1\right|}

\makeatletter

\@ifundefined{textcolor}{}
{%
	\definecolor{BLACK}{gray}{0}
	\definecolor{WHITE}{gray}{1}
	\definecolor{RED}{rgb}{1,0,0}
	\definecolor{GREEN}{rgb}{0,1,0}
	\definecolor{BLUE}{rgb}{0,0,1}
	\definecolor{CYAN}{cmyk}{1,0,0,0}
	\definecolor{MAGENTA}{cmyk}{0,1,0,0}
	\definecolor{YELLOW}{cmyk}{0,0,1,0}
}

\makeatother

\DeclareMathOperator{\Tr}{Tr}

\begin{document}
	
	\title{Tomography in the presence of stray inter-qubit coupling}
	
	\author{Tanay Roy}
	\affiliation{James Franck Institute, University of Chicago, Chicago, Illinois 60637, USA}
	\affiliation{Department of Physics, University of Chicago, Chicago, Illinois 60637, USA}
	
	\author{Ziqian Li}
	\affiliation{James Franck Institute, University of Chicago, Chicago, Illinois 60637, USA}
	\affiliation{Department of Physics, University of Chicago, Chicago, Illinois 60637, USA}
	
	\author{Eliot Kapit}
	\affiliation{Department of Physics, Colorado School of Mines, Golden, CO 80401}
	\affiliation{Department of Physics and Engineering Physics, Tulane University, New Orleans, LA 70118}
	
	\author{David I. Schuster}
	\email{Corresponding author: David.Schuster@uchicago.edu}
	\affiliation{James Franck Institute, University of Chicago, Chicago, Illinois 60637, USA}
	\affiliation{Department of Physics, University of Chicago, Chicago, Illinois 60637, USA}
	\affiliation{Pritzker School of Molecular Engineering, University of Chicago, Chicago, Illinois 60637, USA}
	
	\date{\today}
	
	\begin{abstract}
		Tomography is an indispensable part of quantum computation as it enables diagnosis of a quantum process through state reconstruction. Existing tomographic protocols are based on determining expectation values of various Pauli operators which typically require single-qubit rotations. However, in realistic systems, qubits often develop some form of unavoidable stray coupling making it difficult to manipulate one qubit independent of its partners. Consequently, standard protocols applied to those systems result in unfaithful reproduction of the true quantum state. We have developed a protocol, called coupling compensated tomography, that can correct for errors due to parasitic couplings completely in software and accurately determine the quantum state. We demonstrate the performance of our scheme on a system of two transmon qubits with always-on \textit{ZZ} coupling. Our technique is a generic tomography tool that can be applied to large systems with different types of stray inter-qubit couplings and facilitates the use of arbitrary tomography pulses and even non-orthogonal axes of rotation.
		
	\end{abstract}

	\maketitle
	
	\section{Introduction}
	
	The ability to accurately characterize a quantum state plays a crucial role in benchmarking quantum evolutions and improving quantum control. Quantum state tomography (QST) is used to completely characterize a physical system by reconstructing the density matrix of its state. Any QST scheme involves three steps --- (1) preparation of a large number of identical copies of the quantum state, (2) manipulation of the states specific to a given observable, and (3) projective measurements to determine various coincidence counts. The expectation values for a set of non-commuting observables are then determined from the experimental outcomes and further used to reconstruct the density matrix. One of the most challenging aspects of QST is to precisely manipulate the system so that the measurement operator corresponding to a given observable can be achieved. The difficulty arises from the fact that any real system suffers from miscalibrated pulses (often due to system drifts), finite coherences, readout errors, and qubit crosstalk. Among these, parasitic coupling between qubits is emerging as a major problem, in particular for superconducting circuits, as quantum processors are growing from small-scale \cite{Dicarlo2009TwoQubit_algo, Martinis2010GHZ_W, Martinis2012prime_factor, Siddiqi2018VQE, Eichler2019entanglement, Vijay2020trimon, Pan2017ten_qubit} to intermediate-sized ones \cite{Google2019supremacy, IBM2021Rochester}. While efforts have been made to find ways of suppressing unwanted crosstalk by designing tunable coupler circuits~\cite{Martinis2014tunable_coup, Oliver2018tunable_coup, Houck2019Suppress, Sun2020tunable_coup, Oliver2020tunable_coup}, engineering inter-qubit coupling architecture~\cite{Schuster2015coupler, McKay2020supress_ZZ} and integrating qubits with opposite anharmonicities~\cite{Plourde2020suppression} performing tomography in the presence of stray inter-qubit couplings is still important.
	
	The simplest form of tomography is called \textit{direct inversion} or \textit{linear tomography} where a set of operators spanning the Hilbert space are chosen and then the experimentally determined expectation values of those operators are directly used to compute the density matrix $\rho$~\cite{Schmied2014QST_comparison}. Traditionally, (multi-qubit) Pauli operators are chosen with $\sigma_z$ as the measurement basis. Single qubit rotations about $x$ and $y$ axes are performed to measure $\sigma_y$ and $\sigma_x$ operators. Next coincidence counts are measured from repeated projective measurements to determine the expectation values. However, due to experimental inaccuracies and statistical fluctuations, direct inversion often leads to unphysical density matrices~\cite{White2001tomo2q, Schmied2014QST_comparison, Englert2014QST}. There are well-developed techniques~\cite{Hradil1997MLE, Sacchi1999MLE, Rauch2000MLE, Jezek2001MLE, Lvovsky2007MLE, Shang_2013, Titchener2018QST} to overcome this issue based on maximum-likelihood estimation (MLE) that finds the most probable density matrix compatible with the experimental outcomes by optimizing a likelihood function. Other studies report on using an overcomplete set of measurement operators for MLE to improve the accuracy of the tomographic reconstruction~\cite{Gilchrist2008overcomplete, Senellart2019overcomplete}. 
	
	The most important part of the standard QST protocols is to compute the expectation values of different Pauli operators which is trivial if individual qubits can be rotated independently. But measuring those multi-qubit Pauli operators becomes challenging for a system with non-zero inter-qubit coupling as the application of the tomography pulses (also called pre-rotations) alters a larger part of the Hilbert space. While methods are being investigated for mitigating the effects of parasitic couplings, it might not always be possible or practical to do so. Thus a tomography protocol capable of handling generic multi-qubit systems is needed. We propose and demonstrate a method, called \textit{coupling compensated tomography} (CCT), that can faithfully reconstruct the quantum state of a physical system with arbitrary inter-qubit coupling.
	
	The basic idea behind CCT is to correct the error caused by finite stray couplings in software. This is done by simulating the dynamics of the system during the application of the tomography pulses and determining ideal coincidence counts. Besides the overhead of computing evolution matrices, the complexity of this method is the same as the standard MLE. The correctness of CCT depends on the knowledge of the system Hamiltonian which usually can be determined accurately from the device geometry and experiments.

	\section{Theory}
	\label{sec:theory}
	
	\begin{figure*}[t]
		\centering\includegraphics[width=\textwidth]{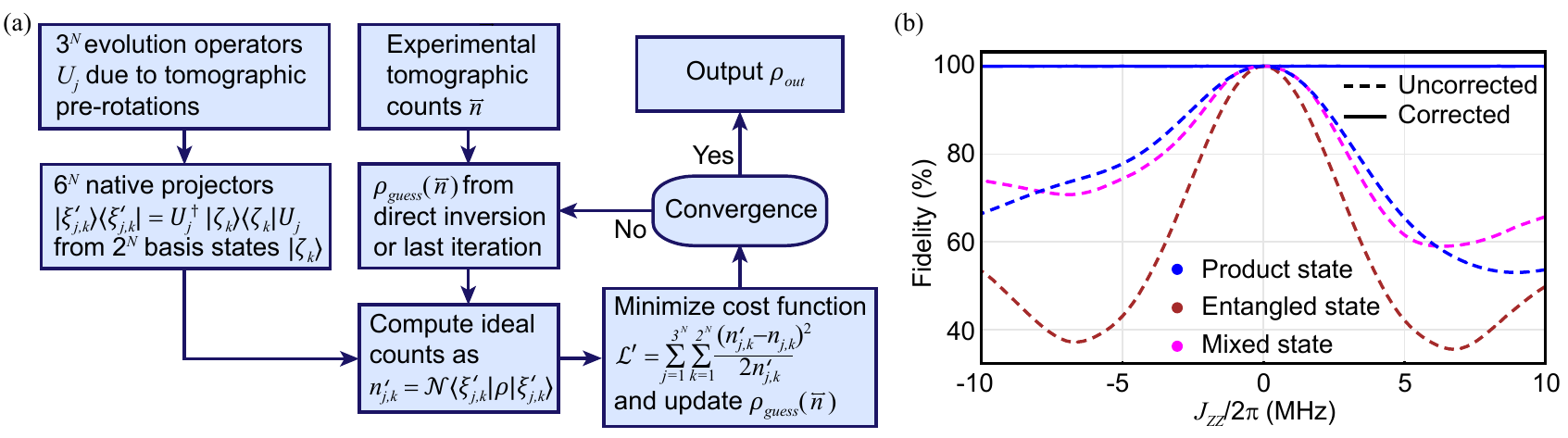}
		\caption{ Coupling compensated tomography scheme and its application on two-qubit system. (a) Flow chart explaining the steps involved in CCT. The central idea is to compute expected coincidence counts due to various pre-rotations using the system Hamiltonian and minimize a likelihood function to reconstruct the density matrix. (b) Comparison of simulated fidelities for a two-qubit system in the presence of inter-qubit \textit{ZZ} coupling. The dashed lines present fidelities obtained with standard maximum likelihood estimation which doesn't incorporate the \textit{ZZ} coupling and the solid lines present the same with CCT as a function of \textit{ZZ} coupling strength. Three representative states (see text for description) --- a product (blue), an entangled (brown), and a mixed (magenta) are chosen to show the effectiveness of CCT which can perfectly reconstruct the original states.}
		\label{fig:fig1}
	\end{figure*}
	
	We consider the following generic Hamiltonian for a system consisting of $N$ qubits
	\begin{equation}
	\label{eq:Hzz}
	\mathcal{H} = H_0 + H_{\rm coup} = \sum_{k=1}^{N} \hbar \omega_{k} \ket{e_k}\bra{e_k} + H_{\rm coup},
	\end{equation}
	where $H_{\rm coup}$ represents an always-on inter-qubit coupling, $\omega_{k}$ is the frequency and $\ket{e_k}$ represents the excited state of the $k$-th qubit. An arbitrary state of the system is represented by a $2^N \times 2^N$ dimensional density matrix $\rho$. One can express $\rho$ as
	\begin{equation}
	\label{eq:rho_expansion}
	\rho = \dfrac{1}{2^N} \sum_{k_1, k_2, \cdots , k_N=0}^{3} r_{k_1, k_2,\cdots, k_N} \hat{\sigma}_{k_1} \otimes \hat{\sigma}_{k_2} \otimes \cdots \otimes \hat{\sigma}_{k_N},
	\end{equation}
	where $r_{k_1, k_2,\cdots, k_N}$ are expectation values (real) corresponding to $4^N$ multi-qubit Pauli operators $\hat{\mu}_m = \hat{\sigma}_{k_1} \otimes \hat{\sigma}_{k_2} \otimes \cdots \otimes \hat{\sigma}_{k_N}$. Here $\sigma_0, \sigma_1, \sigma_2$ and $\sigma_3$ represent identity and three conventional Pauli matrices $\sigma_x, \sigma_y$ and $\sigma_z$ respectively. One can choose the Pauli operators or any other set of $4^N$ orthogonal measurement operators $\hat{\mu}_m = \ket{\xi_m}\bra{\xi_m}$ spanning the Hilbert space according to experimental convenience. Given the total number of repetitions $\mathcal{N}$ for a particular projector, the average number of coincidence counts that will be observed is
	\begin{equation}
	{n'_m} = \mathcal{N}\Tr \left[ \rho \ket{\xi_m}\bra{\xi_m} \right] = \mathcal{N}\bra{\xi_m} \rho \ket{\xi_m}.
	\end{equation}
	The expectation values $r_m$ are then determined from experimentally obtained tomographic counts $\vec{n}$ and Eq.~\eqref{eq:rho_expansion} is used to find the density matrix $\rho_{\rm linear}(\vec{n})$. This method of determining the density matrix, known as linear tomography or direct inversion, often leads to unphysical density matrices~\cite{Schmied2014QST_comparison, White2001tomo2q} and it has become a standard to use MLE to mitigate this issue. In regular MLE, one defines the ``likelihood" function~\cite{White2001tomo2q}
	\begin{equation}
	\label{eq:likelihood_function}
	\mathcal{L} = \sum_{m=1}^{4^N} \frac{\left( \mathcal{N}\bra{\xi_m} \rho \ket{\xi_m} -n_m\right)^2} {\mathcal{N}\bra{\xi_m} \rho \ket{\xi_m}},
	\end{equation}
	and minimizes it to obtain the physical density matrix $\rho_{\rm MLE}$ that best describes the state of the system. It is common to use $\rho_{\rm linear}(\vec{n})$ as an initial guess for the minimization process.
	
	While certain architectures allow joint measurements~\cite{Wallraff2009JointRdt}, superconducting qubits usually have individual readouts~\cite{Devoret2013outlook} enabling measurements of single-qubit operators. A projective measurement in such a system leads to $2^N$ $N$-coincidence counts $n_{k_1,k_2,\cdots,k_N},$ with $k_j \in \{0,1\}$ corresponding to a projector $\ket{\zeta_{k_1,k_2,\cdots,k_N}}\bra{\zeta_{k_1,k_2,\cdots,k_N}}$ where $\ket{\zeta_{k_1,k_2,\cdots,k_N}}$ is one of the basis states (usually along the $z$-axis). Measurements along any other direction requires pre-rotations of the qubits achieved by the drive Hamiltonian
	\begin{equation}
	\label{eq:Hdrive}
	H = \sum_{k=1}^{N} A_k(t) \sin(\omega_k t - \phi_k) \sigma_{x_k},
	\end{equation}
	where $A_k(t)$ and $\phi_k$ are the drive amplitudes and phases respectively. Since it is convenient to measure along the six cardinal points of individual Bloch spheres, to perform a full tomography, a set of $3^N$ projective measurements is performed with appropriate pre-rotations. The tomographic counts $\vec{n}$ (having $6^N$ elements) from all measurements are used to compute the expectation values $r_m$ corresponding to the Pauli operators $\hat{\mu}_m$ and then the density matrix is reconstructed by minimizing the likelihood function
	\begin{equation}
	\label{eq:likelihood_2}
	\mathcal{L} = \sum_{m=1}^{4^N} \frac{\left( \bra{\xi_m} \rho \ket{\xi_m} -r_m\right)^2} {2\bra{\xi_m} \rho \ket{\xi_m}}.
	\end{equation}
	
	Traditionally, the drives in Eq.~\eqref{eq:Hdrive} perform $\pi/2$ rotations of each qubit around $x$- or $y$-axis to enable projections along the Cartesian axes of the Bloch spheres. However, this manipulation requires effective rotation of one qubit independent of its partners which becomes difficult to achieve in the presence of inter-qubit coupling. We overcome this hurdle by employing measurement operators that are native to the system and using experimental tomographic counts in the ``likelhood" function, instead of expectation values.
	
	Fig.~\ref{fig:fig1}(a) depicts the flow diagram of CCT. The central idea is to compute ideal tomographic counts $\overleftrightarrow{n}'$ accounting for the effect of $H_{\rm coup}$. We first calculate $3^N$ evolution operators corresponding to the pre-rotations applied to the system as
	\begin{equation}
	\label{eq:evol_op}
	U_j(t_0,t) = \mathcal{T}e^{-i\displaystyle\int_{t_0}^t \left(H_0+H_{\rm coup} + H_j(t) \right)dt},
	\end{equation}
	where $t_0$ is the beginning of the pulse being applied and $\mathcal{T}$ represents the \textit{time-ordering} operator. Obtaining analytic expressions (after applying rotating wave approximation) for $U_j$ is possible if $H_{\rm coup}$ is diagonal in the computational basis and $A_k(t)$ has simple pulse shapes. One such example would be 
	$H_{\rm coup}=J_{j,k}\ket{e_ke_j}\bra{e_ke_j}$ describing cross-Kerr (also known as \textit{ZZ}) coupling, a common form of parasitic interaction~\cite{Steffen2011CRgate, Google2020fsim, Martinis2019diabatic, Wallraff2020Cphase}, with rectangular tomography pulses. However, in general, one will need numerical techniques to evaluate Eq.~\eqref{eq:evol_op}. Note that one of the evolution operators is always identity and thus, in practice, $3^N-1$ evolution operators need to be determined. Next, $6^N$ modified projectors $\hat{\mu}'_{j,k}=\ket{\xi'_{j,k}}\bra{\xi'_{j,k}}$ are calculated from
	\begin{equation}
	\label{eq:xi_prime}
	\ket{\xi'_{j,k}} = U_j^\dagger \ket{\zeta_k}, \ \left\{
	\begin{array}{cc}
	j &\in \{1,\cdots,3^N\}  \\
	k & \in \{1,\cdots, 2^N\} 
	\end{array}
	\right. .
	\end{equation}
	When $H_{\rm coup}=0$, the projectors $\hat{\mu}'_{j,k}$ essentially coincide with the cardinal points of the Bloch spheres. Consequently, the goal becomes minimization of the modified likelihood function
	\begin{equation}
	\label{eq:likelihood_3}
	\mathcal{L}' = \sum_{j=1}^{3^N} \sum_{k=1}^{2^N} \frac{\left( \mathcal{N}\bra{\xi'_{j,k}} \rho \ket{\xi'_{j,k}} -n_{j,k}\right)^2} {2 \mathcal{N} \bra{\xi'_{j,k}} \rho \ket{\xi'_{j,k}}},
	\end{equation}
	where $n'_{j,k}=\mathcal{N}\bra{\xi'_{j,k}} \rho \ket{\xi'_{j,k}}$ are the expected coincidence counts in the presence of $H_{\rm coup}$. Note that we have considered an overcomplete set of measurements to improve the accuracy of the state reconstruction~\cite{Gilchrist2008overcomplete, Senellart2019overcomplete}, but one can choose any $4^N$ projectors spanning the Hilbert space.
	
	As an example, we demonstrate the effectiveness of our method in the presence of one of the most commonly occurring coupling --- cross-Kerr~\cite{Gambetta2016Tuneup, Google2020fsim, Martinis2019diabatic, Wallraff2020Cphase}. We simulate a two-qubit system in the presence of the coupling Hamiltonian, 
	\begin{equation}
	H_{\rm coup} = J_{zz} \ket{e_2 e_1}\bra{e_2 e_1}.
	\end{equation}
	This form of coupling makes the transition frequency of one qubit dependent on the state of its partner and is an unwanted feature~\cite{Steffen2011CRgate, Eichler2019entanglement, Sun2020tunable_coup} (when always-on) for most systems aimed for quantum computation. Fig.~\ref{fig:fig1}(b) plots the simulated fidelity of the tomographic reconstruction for three randomly chosen initial states --- (1) a product state (purple lines): $\ket{\psi_p} = (\ket{g}+\ket{e})^{\otimes 2}/2$, (2) an entangled state (brown lines): $\ket{\psi_e} = (\ket{gg} + \ket{ee})/\sqrt{2}$, and (3) a mixed state (magenta lines): $0.8\ket{\psi_p}\bra{\psi_p} + 0.2\ket{\psi_e}\bra{\psi_e}$ as a function of cross-Kerr coupling strength $J_{zz}$. The dashed lines show that the states are not correctly reproduced for non-zero $J_{zz}$ when the regular tomography is used, whereas solid lines show that CCT completely recovers the correct state.
	
	
	\begin{figure*}[t]
		\centering\includegraphics[width=\textwidth]{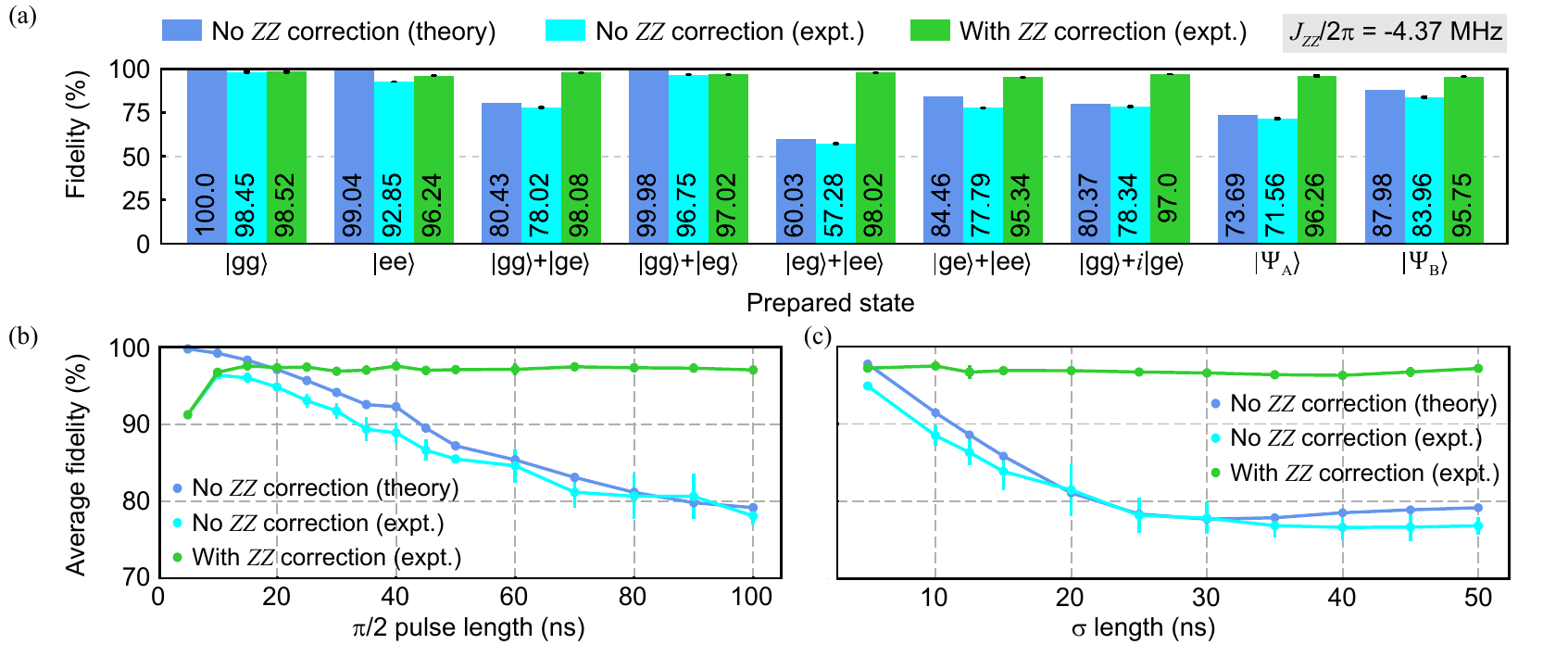}
		\caption{Comparison of fidelities. (a) Bar graph showing fidelities for different initial states. Fidelities obtained theoretically using regular MLE (blue), experimentally using regular MLE (cyan), and experimentally using CCT (green) are plotted for nine different initial states (normalization coefficients are not shown for brevity). All experimental fidelities are average of four tomographic reconstructions each with 5000 projective measurements for every pre-rotation using 50 ns long rectangular $\frac{\pi}{2}$ pulses and corrected for measurement error (see Supplementary section XII). (b) Comparison of average fidelities obtained with 12 different initial states as a function of Rabi drive strength. Rectangular pulses are used with $\frac{\pi}{2}$ pulse length being tuned from 5 ns to 100 ns. Experimental fidelities for the Rabi rate $\ge 25$ MHz (first two data points) are lower due to leakage of excitation to higher states and imperfect $\frac{\pi}{2}$ gate calibration. (c) Comparison of average fidelities with Gaussian tomographic pulses as a function of $\sigma$. Both $\frac{\pi}{2}$ and $\pi$ pulses are $4\sigma$ long.}
		\label{fig:fig2}
	\end{figure*}

	\section{Experiment}
	\label{sec:expt}
	
	\begin{table}[b]
		\centering
		\begin{tabular}{||c|c|c|c|c|c||}
			\hline
			Qubit & $f$ (GHz) & $\alpha/2\pi$ (GHz) & \textrm{$T_1 (\mu \rm s)$} & \textrm{$T_{\rm R} (\mu \rm s)$} & \textrm{$T_{\rm e} (\mu s)$} \\
			\hline
			\textrm{q1} & 3.49428 & $-0.157$ & 31.5 & 30.4 & 35.3 \\ 
			\hline
			\textrm{q2} & 4.23200 & $-0.188$ & 12.2 & 16.0 & 17.7\\ 
			\hline
		\end{tabular}
		\caption{Frequencies $(f)$, anharmonicities ($\alpha$), relaxation times $T_1$, Ramsey times $T_R$ and echo times ($T_e$) for the two qubits.}
		\label{table:qubit_param}
	\end{table}
	
	Our device consists of two transmons~\cite{koch2007transmon} with individual readout resonators. A superconducting quantum interference device (SQUID) is used as a coupler between the two transmons. The details of the device are presented in Supplementary section IV. Stray capacitive and inductive couplings lead to cross-Kerr interaction between the qubits. We extract the cross-Kerr strength $J_{zz}$ by performing the Ramsey experiment on one qubit when the other qubit is in its ground or excited state. Various coherence parameters of the qubits are shown in Table~\ref{table:qubit_param}.

	\subsection{Rectangular pulses}
	\label{square}
	We first consider the simplest form of tomography pulses, namely, the rectangular pulses which are applied for a period $T$ with a constant drive strength $A=\frac{\pi}{2T}$. If a microwave drive is applied to one of the qubits at a time, under rotating wave approximation (RWA), one can obtain analytic expressions for the evolution of arbitrary two-qubit states (see Supplementary section VII). These evolution matrices are used to compute ideal tomographic counts for a given state. We always apply a drive on qubit 1 followed by qubit 2. 
	
	We choose a Rabi rate of 5 MHz when the partner qubit is in its ground state so that the $\pi/2$ pulses are 50 ns long. At our operating point, cross-Kerr strength $J_{zz}$ is $-4.37$ MHz. Here the negative sign indicates that the frequency of qubit 1(2) decreases when qubit 2(1) is excited to $\ket{e}$. Fig.~\ref{fig:fig2}(a) shows a comparison of fidelities for different prepared states, where the blue bars show theoretically expected fidelities when no \textit{ZZ} correction is applied, cyan bars show experimentally obtained fidelities without \textit{ZZ} correction being applied and green bars represent experimentally obtained fidelities with CCT. Our experimental tomographic counts $\vec{n}$ are corrected for readout error (see Supplementary section XII). The first seven product states are exactly (up to experimental accuracy) prepared by applying combinations of $\pi$ and $\pi/2$ pulses at appropriate frequencies. For example, the state $(\ket{eg} + \ket{ee})/\sqrt{2}$ is prepared by applying a $\pi_y$ pulse on qubit 1 at $f_{q1,q2=\ket{g}}=3.49428$ GHz followed by a $(\pi/2)_y$ pulse on qubit 2 at $f_{q2,q1=\ket{e}}=4.22763$ GHz. Here $\ket{\Psi_{\rm A}}$ is obtained by first preparing $(\ket{gg}+\ket{eg})/\sqrt{2}$ followed by applying a $(\pi/2)_y$ pulse on qubit 2 at $f_{q2,q1=\ket{g}}=4.23200$ GHz. The state $\ket{\Psi_{\rm B}}$ is obtained by waiting for $\pi/J_{zz}$ after preparing $\ket{\Psi_{\rm A}}$. Both $\ket{\Psi_{\rm A}}$ and $\ket{\Psi_{\rm B}}$ are entangled states due to the presence of finite $J_{zz}$ (see Supplementary section VIII for explicit expressions). Similar comparison for $J_{zz}=-4.90$ MHz and $J_{zz}=-5.66$ MHz are shown in Supplementary section IX. Note that the fidelities are computed with respect to the theoretically expected state and we have used ``L-BFGS-B" optimization algorithm to minimize the likelihood function in Eq.~\eqref{eq:likelihood_3}. It is clearly visible that the uncompensated cases are unable to reconstruct the correct states and match very well with theoretically expected fidelities, whereas, CCT almost fully recovers the state by successfully correcting the error due to \textit{ZZ} coupling.
	
	Next, we show that this scheme works with different drive strengths. Fig.~\ref{fig:fig2}(b) plots the average fidelity as a function of $\pi/2$-pulse lengths (identical for both qubits) where the average is obtained for the states showed in Fig.~\ref{fig:fig2}(a) along with $\ket{ge}, \ket{eg}$ and $(\ket{gg}+i\ket{eg})/\sqrt{2}$. For progressively slower pulses the effect of $J_{zz}$ becomes stronger and thus uncorrected fidelities drop for longer pulses. The green dots clearly show that the correction scheme is independent of drive strengths with average fidelity being above 96\% until the pulses are made very fast. Both the uncorrected (cyan line) and corrected (green line) show poor fidelities for drive strengths $\ge 25$ MHz due to leakage into higher states and imperfect ${\pi}/{2}$ gate calibration.

	\begin{figure}[t]
		\centering\includegraphics[width=\columnwidth]{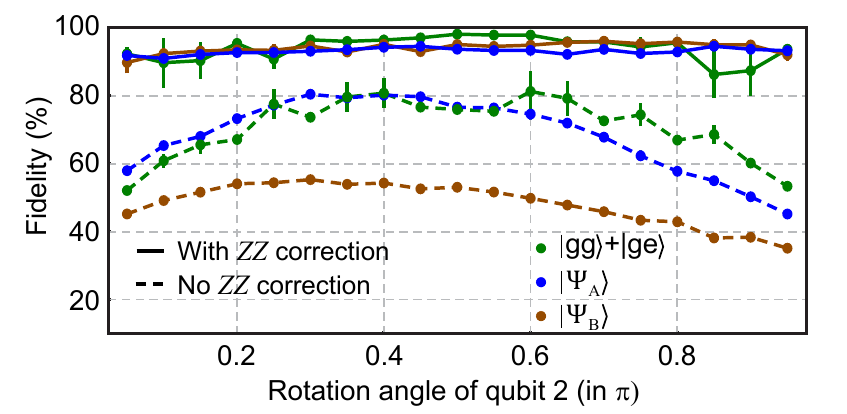}
		\caption{ CCT with non-$\frac{\pi}{2}$ pulses. Fidelities as a function of rotation angle of qubit 2 during tomography when the same for qubit 1 is set to $0.35\pi$. Dash lines represent fidelities with regular MLE while solid lines represent the same with \textit{ZZ} correction through CCT. Red: $(\ket{gg)+\ket{ge}})/\sqrt{2}$, Blue: $\ket{\Psi_{\rm A}}$, Cyan: $\ket{\Psi_{\rm B}}$. Each data point is an average of 4 measurement with error bars showing corresponding standard deviations.}
		\label{fig:fig3}
	\end{figure}
	
	\subsection{Other pulse shapes}
	\label{other_waveforms}
	Rectangular pulses are not always suitable due to its large bandwidth. Other common waveforms include Gaussian~\cite{BAUER1984442, Chuang2003accurate_control}, Gaussian filtered flat-top~\cite{Schuster2017multimode2d, Michielsen2017Gate-error}, DRAG (derivative removal via adiabatic
	gate)~\cite{Wilhelm2009DRAG, Gambetta2016Tuneup}, SWIPHT (speeding up waveforms by inducing phases to harmful
	transitions)~\cite{Barnes2015SWIPHT} and more recently optimal-control pulses~\cite{Ginossar2017optimal_control, Filipp2021optimal_control} to avoid leakage to non-computational subspace. While analytical expressions for non-rectangular pulses are difficult to calculate, the evolution matrices can always be computed numerically. We use QuTip~\cite{qutip1, qutip2} to calculate the evolution matrices. For the purpose of demonstration we choose Gaussian pulses with a cutoff of $\pm2\sigma$. In Fig.~\ref{fig:fig2}(c) average fidelity for the three different cases are shown as function of $\sigma$. The uncorrected fidelities (cyan points) obtained experimentally match pretty well with the theoretical values (blue points) and the corrected fidelity (green points) is $> 95\%$ for the whole range. This is a clear indication that CCT is robust against the choice of waveform profiles.
	
	
	\subsection{Non-orthogonal measurement axes}
	\label{non_ortho}
	
	Traditional tomography protocols utilize correlators defined by Pauli operators and hence depend on accurate calibration of $\pi/2$ rotations about orthogonal axes. CCT, on the other hand, empowers use of non-orthogonal measurement axes. There are two implications of this aspect --- use of non-$\pi/2$ pulses for pre-rotations and rotations about non-orthogonal axes.
	
	We demonstrate the first feature in Fig.~\ref{fig:fig3}. Here, we sweep the rotation angle from $0.05\pi$ to $0.95\pi$ for qubit 1 while using $0.35\pi$ rotation for qubit 2 during application of tomography pulses (rectangular). The dashed lines represent fidelities without \textit{ZZ} correction for the three states $(\ket{gg}+\ket{ge})/\sqrt{2}$ (green), $\ket{\Psi_{\rm A}}$ (blue) and $\ket{\Psi_{\rm B}}$ (brown) and the solid lines represent corresponding fidelities with \textit{ZZ} correction applied through CCT. It is clearly visible that the performance of CCT is always superior.
	
\begin{figure}[t]
		\centering
		\includegraphics[width=\columnwidth]{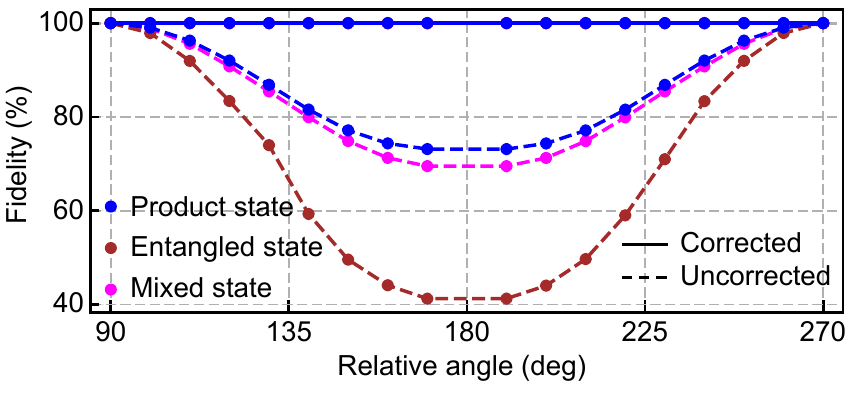}
		\caption{CCT with non-orthogonal axes of rotation. Simulated fidelity of reconstructed states when the relative angle of the second axis of rotation is swept from $90^\circ$ to $270^\circ$. The blue, brown and magenta colored points represent a product state: $\ket{\psi_p} = (\ket{g}+\ket{e})(\ket{g}+\ket{e})/2$, an entangled state: $\ket{\psi_e} = (\ket{gg} + \ket{ee})/\sqrt{2}$, and a mixed state: $0.8\ket{\psi_p}\bra{\psi_p} + 0.2\ket{\psi_e}\bra{\psi_e}$. The data points at $180^\circ$ are removed since colinear axes of rotation do not provide enough information to reconstruct the state.}
		\label{fig:non_ortho}
	\end{figure}

	Another useful advantage of CCT is the ability to use non-orthogonal axes of rotation for the pre-rotations. Fig.~\ref{fig:non_ortho} shows simulation of state reconstruction for an uncoupled two-qubit system ($J_{zz}=0$ in Eq.~\eqref{eq:Hzz}). The same states as in Fig.~\ref{fig:fig1}(b) are considered for demonstration. As the relative angle between the axes of rotation deviates from standard $90^\circ$, regular tomography considering Pauli operators produces progressively inaccurate states, whereas, CCT correctly determines the prepared state.

	\section{Conclusion}
	In conclusion, we have presented a scheme to perform tomography of a multi-qubit system in the presence of inter-qubit coupling. The coupling compensated tomography utilizes measurement operators that are natural to the system and can be regarded as a generalization to the standard methods that consider orthogonal projectors. At the core of CCT lies the computation of ideal measurement statistics after the application of different tomography pulses and finding the most probable state that explains the experimental observations through an optimization routine. The efficacy depends on the accuracy of the Hamiltonian describing the physical system.  The most important feature of CCT is that it can compensate for the effect of stray inter-qubit couplings completely in software without increasing operational complexity. The only challenging aspect would be the (one-time) computation of the $3^N$ evolution matrices as the system size grows. Nevertheless, larger systems with certain types of nearest-neighbor (or second nearest-neighbor) coupling (e.g. \textit{ZZ}) that allow decomposition of the Hilbert space into separable subsystems can be simulated efficiently. Besides this overhead, the computational complexity of CCT is identical to the standard MLE-based protocols.
	
	We have performed an experimental demonstration on a two-qubit system with always-on cross-Kerr coupling for a wide variety of initial states and tomography pulses. While CCT is capable of perfect reconstruction of states, our results are mainly limited by qubit relaxation and imperfect gate calibration. Further, our method is not limited to any particular type of inter-qubit interaction or number of qubits. Two important features of CCT are the ability to work with non-$\pi/2$ pulses and non-orthogonal rotation axes. Employing sub-$\pi/2$ tomography pulses will be beneficial for systems with not well-calibrated stray couplings while tomography with unconventional rotation axes will help platforms where native orthogonal rotational axes unavailable~\cite{Schuster2021Fluxonium, Gossard2010non_ortho}. We believe CCT, being a versatile tomography protocol, is poised to be a useful characterization tool for various quantum computing platforms.

	\begin{acknowledgments} 
		This work was supported by the Army Research Office under Grant No. W911NF-18-1-0125 and the National Science Foundation Grant No. PHY-1653820. This work was partially supported by the University of Chicago Materials Research Science and Engineering Center, which is funded by the National Science Foundation under award number DMR-1420709. Devices are fabricated in the Pritzker Nanofabrication Facility at the University of Chicago, which receives support from Soft and Hybrid Nanotechnology Experimental (SHyNE) Resource (NSF ECCS-1542205), a node of the National Science Foundation’s National Nanotechnology Coordinated Infrastructure. 
	\end{acknowledgments} 
	
	\input{main_ZZ_arxiv.bbl}

	\clearpage
	\setcounter{figure}{0}
	\setcounter{table}{0}
	\setcounter{equation}{0}
	\setcounter{section}{0}
	\global\long\def\theequation{S\arabic{equation}}
	\global\long\def\thefigure{S\arabic{figure}}
	\global\long\def\thetable{S\arabic{table}}
	\renewcommand{\theHtable}{Supplement.\thetable}
	\renewcommand{\theHfigure}{Supplement.\thefigure}
	\renewcommand{\theHequation}{Supplement.\theequation}
	\onecolumngrid
	\begin{center}
		{\bf \large Tomography in the presence of stray inter-qubit coupling: Supplementary Information}
		\newline
		Tanay Roy \textit{et al.}
	\end{center}
	
	\section{Measurement setup}
	\label{supp:extra}
	
\begin{figure}[h]
		\includegraphics[width=\textwidth]{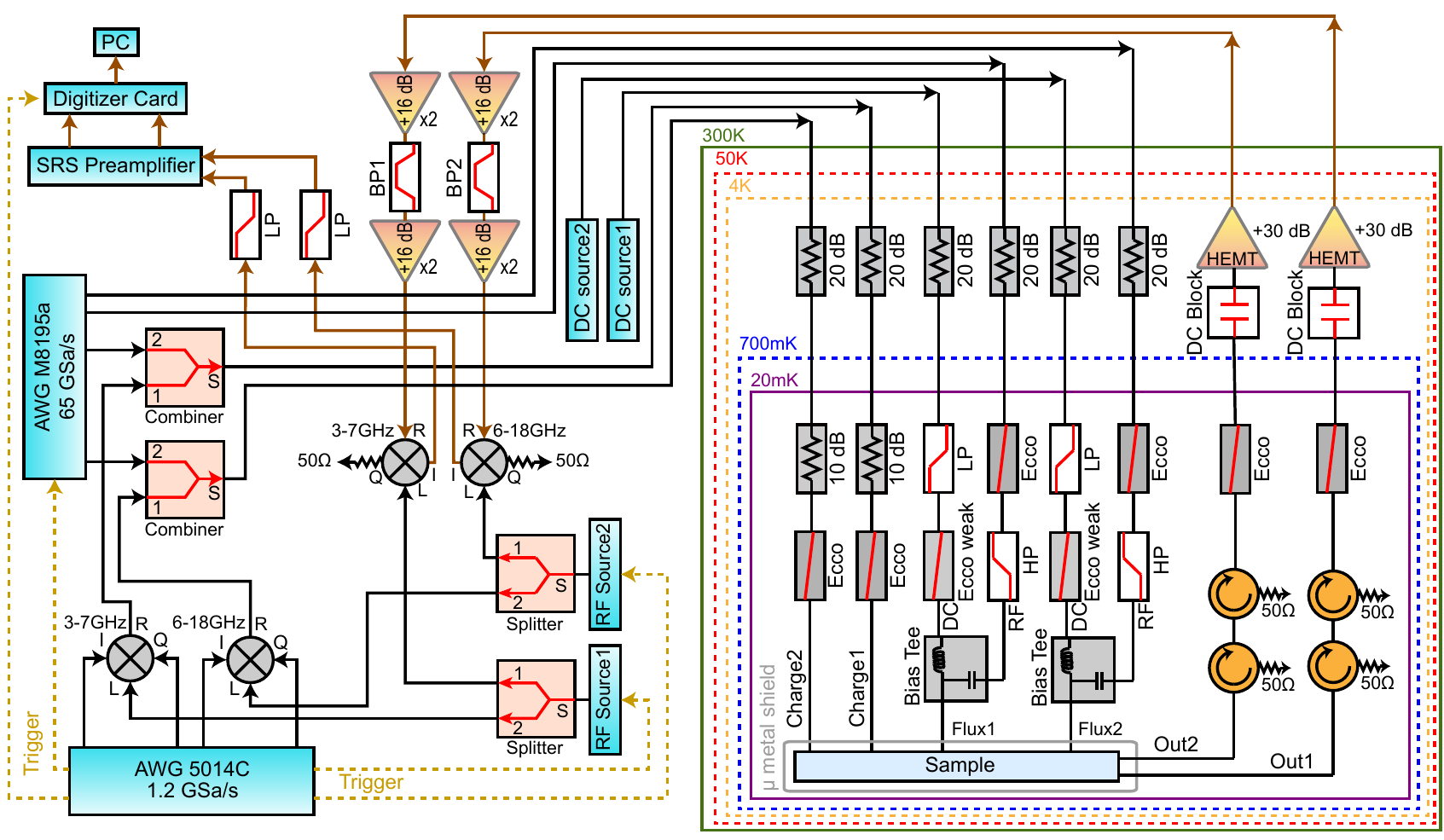}
		\caption{Detailed measurement setup with room temperature and cryogenic circuitry.}
		\label{fig:setup}
		\centering    
	\end{figure}

	Fig.~\ref{fig:setup} shows the detailed measurement setup. The device is mounted inside a light-tight cylindrical copper can having a bilayer $\mu$-metal shield on outside and measured inside a dilution refrigerator with a base temperature of 20 mK. The readout pulses are generated by modulating CW tones from two RF sources (PSG-E8257D) using an arbitrary waveform generator (AWG). The AWG (Tektronix 5014C) running at 1.2 GSa/s also acts as a master trigger for the rest of the equipment. The pulses for the qubits are directly synthesized using a second AWG (Keysight M8195a) with a sampling rate of 16 GSa/a. The readout and qubit pulses are combined before entering the fridge (through charge lines). Two current sources (Yokogawa GS200) are used to apply DC fluxes to the loops. All input signals are attenuated by 20-dB attenuators at the 4-K stage. The charge lines are further passed through 10-dB attenuators and lossy Ecoosorb\textsuperscript{\textregistered} filters at the base plate. Low pass filters (with 1.9 MHz cutoff) and weak Ecoosorb filters are inserted on the DC flux lines at the base stage. We have the capability of applying RF flux pulses but are not utilized in this project. The transmitted signals are amplified using commercial HEMT (LNF) amplifiers at the 4-K stage after passing through circulators, weak Eccorsorb filters and DC blocks. The output signals are further amplified at the room temperature after appropriate filtering (bandpass) before being demodulated using IQ mixers (Marki). The demodulated homodyne signals are low-pass filtered and pre-amplified followed by digitization (using Alazar ATS 9870) at 1 GSa/s sampling rate. The digitzed signals are stored and analyzed in a computer.

	\section{Device fabrication}
	\label{supp:fab}
	The device is fabricated on a 430-$\mu$m-thick C-plane sapphire wafer with Niobium as the base layer. The wafer is first annealed at 1200$^{\circ}$ C followed by deposition of 75 nm Niobium through electron-beam evaporation. The large features (resonators, capacitor pads and input-output lines) are made using photolithography and reactive ion etch (RIE) at wafer scale. The wafer is spin-coated with about 600 nm thick AZ MiR 703 (positive) photoresist which is exposed with 375 nm laser using a Heidelberg MLA150 Direct Writer. The exposed photoresist is developed with AZ 300 MIR developer followed by RIE performed using a PlasmaTherm ICP Fluorine etching tool. We fabricate the Dolan bridge \cite{Dolan1977JJ} style Josephson junctions whose masks are created by electron-beam lithography in a Raith EBPG5000 Pluse writer. The e-beam bilayer consists of 500 nm thick MMA EL11 (bottom layer) and 500 nm thick 950 PMMA A7 resist (top layer). The e-beam resists, exposed with 100 kV electron beam, are developed with a solution of 3:1 IPA:water for 90 seconds. Aluminum is evaporated on the wafer at angles $\pm20^\circ$ inside a Plassys MEB550S electron beam evaporator with intermediate oxidation for 12 minutes at 50 mbar (using Ar:O$_2 = 85:15$). The wafer is then diced into $7\times7$ mm chips, followed by liftoff. The chips are mounted on copper printed circuit boards and wire-bonded to make electrical connections.

	\section{Device details}
	\label{supp:device}
	
	\begin{figure}[b]
		\includegraphics[width=\textwidth]{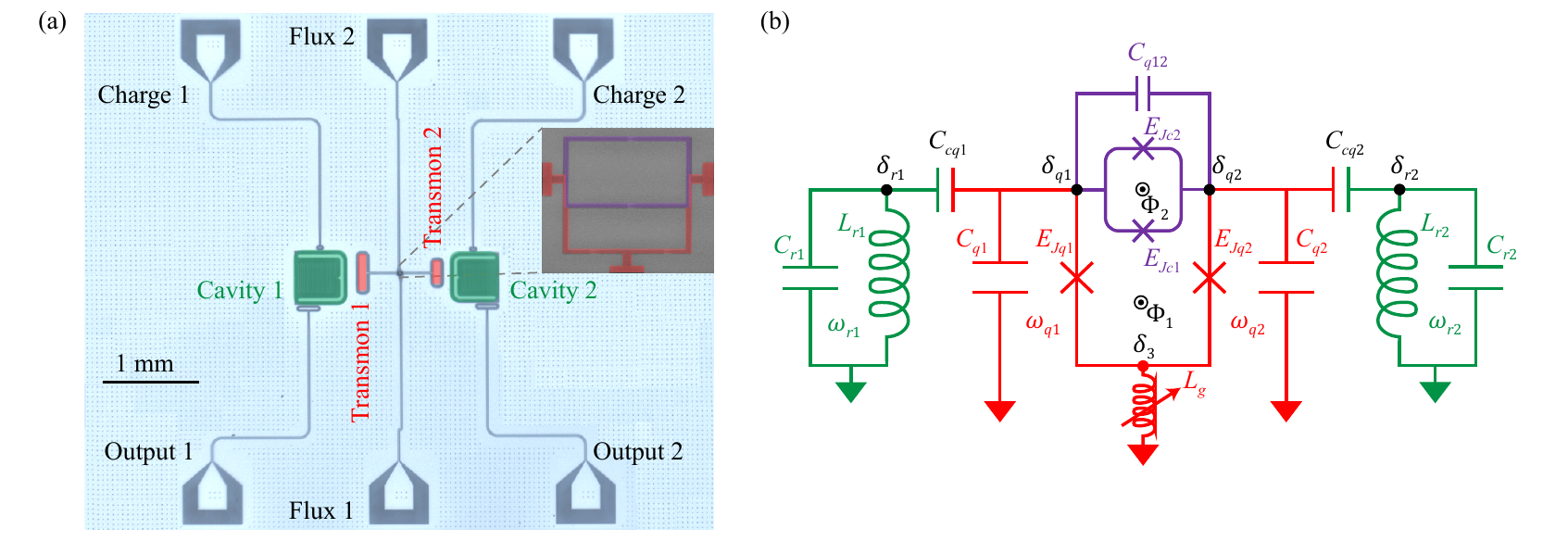}
		\caption{(a) A false-colored optical image of the device. The red and green sections are the transmons and readout resonators respectively. Inset shows a scanning electron micrograph of the coupler loop (purple) and the two Josephson junctions for the qubits forming main loop (red). (d) Schematic diagram. The magnetic fluxes threading the main loop and coupler loop are denoted by $\Phi_1$ and $\Phi_2$ respectively.}
		\label{fig:device}
		\centering    
	\end{figure}
	
	A false-color image of the device is shown in Fig.~\ref{fig:device}(a). It consists of two transmons~\cite{koch2007transmon} (red) coupled through a superconducting quantum interference device (SQUID) as pictured in the inset (purple). The bottom arm of the SQUID together with the junctions of the qubits form the main loop of the device. Each transmon is coupled to an LC resonator (green) for individual readout. This design was originally conceived for the realization of very small logical qubit (VSLQ) and the two flux loops with proper biasing enable relevant interactions~\cite{Kapit2016VSLQ}. A schematic diagram of the device is shown in Fig.~\ref{fig:device}(b). The magnetic fluxes through the main and squid loops are denoted by $\Phi_1$ and $\Phi_2$ respectively.
	
	We operate at $\Phi_2 = \Phi_0/2$ and $\Phi_1=0$, where $\Phi_0$ is the magnetic flux quantum. At this biasing, the squid loop ideally gains infinite inductance and consequently the two qubits become decoupled. However, due to asymmetry in the junctions of the squid loop (because of fabrication uncertainties) and non-zero inductance ($L_g$) of the main loop to the ground, finite inter-qubit \textit{ZZ} coupling develops. We extract the cross-Kerr strength $J_{zz}$ by determining the frequency on one qubit when the other qubit is in its ground or excited state. Ramsey experiments are used to find the qubit frequencies accurately. Further, we have a small squid loop in the shunt to the ground (not visible in Fig.~\ref{fig:device}(a)) to make $L_g$ tunable which in turn allows tuning of $J_{zz}$ (see Supplementary section~\ref{supp:hamil} for details). The $\ket{g} \leftrightarrow \ket{e}$ transitions for the two qubits are $f_{q1,q2=\ket{g}}=3.49428$ GHz and $f_{q2,q1=\ket{g}}=4.23200$ GHz when the partner qubits are in their ground states. The corresponding readout resonators have frequencies $f_{r1}=6.97792$ GHz and $f_{r2}=8.13644$ GHz respectively.
	
	\section{System Hamiltonian}
	\label{supp:hamil}
	
	Fig.~\ref{fig:device}(b) shows the detailed circuit diagram of our device. The inductances, capacitances, Josephson energies, superconducting order parameters and corresponding conjugate charge parameters are denoted by $L_i$s, $E_{Ji}$s, $C_i$s, $\delta_i$s and $Q_i$s respectively. We use the label $q1(2)$ to represent parameters associated with qubit 1(2) and similarly for the two resonators. We first construct the capacitance matrix (considering the four nodes marked black in Fig.~\ref{fig:device}(b))
	\begin{align}
	\bm{C_{0}}=\left[
	\begin{array}{cccc}
	C_{r1}+C_{cq1} & 0 & -C_{cq1} & 0 \\
	0 & C_{r2}+C_{cq2} & 0 & -C_{cq2} \\
	-C_{cq1} & 0  & C_{q1}+C_{cq1}+C_{q12} & -C_{q12} \\
	0 & -C_{cq2}  & -C_{q12} & C_{q2}+C_{cq2}+C_{q12} \\
	\end{array}
	\right].
	\label{eq:cap_mat}
	\end{align}
	Note that here we introduced a capacitance $(C_{q12})$ between the qubits arising mainly due to the self-capacitances of the coupler junctions. Defining the node-flux vector as $\vec{\varphi}^T=(\varphi_1, \varphi_2, \varphi_{q1}, \varphi_{q2})$, we can express the Lagrangian and Hamiltonian of the system as
	\begin{align}
	L &= \frac{1}{2}\vec{\dot{\varphi}}^T \bm{C_0} \vec{\dot{\varphi}}-K, \\
	H &= \frac{1}{2}\vec{Q}^T\bm{C_{0}}^{-1}\vec{Q}+K,
	\end{align}
	where
	\begin{align}
	\label{eq:kinetic}
	K &= \frac{\varphi_1^2}{2L_{r1}} +\frac{\varphi_2^2}{2L_{r2}}+\frac{\varphi_3^2}{2L_g} + U, \\
	\label{eq:J_energy}
	U &= -E_{Jq1}\cos(\delta_{q1}-\delta_{3}) -E_{Jq2}\cos(\delta_{q2}-\delta_{3}) -E_{Jc1}\cos\left(\delta_{q2}-\delta_{q1}+2\pi\frac{\Phi_{1}}{\Phi_0}\right) -E_{Jc2}\cos\left(\delta_{q2}-\delta_{q1}+2\pi\frac{\Phi_{1}+\Phi_{2}}{\Phi_0}\right).
	\end{align}
	Here the first three terms of $K$ are the kinetic energy terms due to the inductors and $U$ represent Josephson energies from the four Josephson junctions. The elements of the charge vector $\vec{Q}^T=(Q_1, Q_2, Q_{q1}, Q_{q2})$ is defined as $Q_i=\frac{\partial L}{\partial \varphi_i}$ and the superconducting order parameters $\delta_i$s are related to the node-fluxes as $\delta_i=2\pi \varphi_i/\Phi_0$.
	
	Disregarding higher order terms of $\varphi_3$ in Eq.~\ref{eq:kinetic}, $\varphi_3$ can be regarded as a non-dynamical variable. To eliminate it, we consider the linearized version of the kinetic energy
	\begin{align}
	K_{\rm lin} = \frac{\varphi_{r1}^2}{2L_{r1}}  +\frac{\varphi_{r2}^2}{2L_{r2}}+\frac{\varphi_3^2}{2L_g}+\frac{\left(\varphi_{q1}-\varphi_3\right)^2}{2L_{Jq1}}+\frac{\left(\varphi_{q2}-\varphi_3\right)^2}{2L_{Jq2}},
	\end{align}
	where $L_{Jqi}=\left(\frac{\Phi_0}{2\pi}\right)^2\frac{1}{E_{Jqi}}$. Next we minimize the Hamiltonian with respect to $\varphi_3$ by setting $\frac{\partial{H}}{\partial{\varphi_3}} =0$ which results in
	\begin{equation}
	\varphi_3 =\frac{L_g}{{L_x}^2}\left(L_{J2}\varphi_{q1}+L_{J1}\varphi_{q2}\right), \ \ 
	{L_x}^2 = L_{Jq1}L_{Jq2}+(L_{Jq1}+L_{Jq2})L_{g}.
	\label{eq:eliminate_phi3}
	\end{equation}
	Plugging Eq.~\ref{eq:eliminate_phi3} back into Eq.~\eqref{eq:J_energy}, we obtain the modified Josephson energy
	\begin{multline}
	U' = -E_{Jq1}\cos\left(x_1\delta_{q1}-y_1\delta_{q2}\right)-E_{Jq2}\cos\left(x_2\delta_{q2}-y_2\delta_{q1}\right) \\
	- E_{Jc1}\cos\left(\delta_{q2}-\delta_{q1}-2\pi\frac{\Phi_{1}}{\Phi_0}\right)+E_{Jc2}\cos\left(\delta_{q2}-\delta_{q1}+2\pi\frac{\Phi_{1}+\Phi_{2}}{\Phi0}\right),
	\label{eq:U_corrected}
	\end{multline}
	where
	\begin{align}
	\left\{ 
	\begin{array}{cc}
	x_1  = &(L_{Jq1}L_{Jq2}+L_{Jq1}L_g)/{L_x^2}, \notag\\
	x_2  = &(L_{Jq1}L_{Jq2}+L_{Jq2}L_g)/{L_x^2}, \notag\\
	y_1  = &{L_{Jq1}L_g}/{L_x^2}, \notag\\
	y_2  = &{L_{Jq2}L_g}/{L_x^2}. \notag\\
	\end{array}
	\right.
	\end{align}
	In order to perform circuit quantization, we find the classical equilibrium point $(\overline{\delta_{q1}}, \overline{\delta_{q2}})$, which satisfies
	\begin{align}
	\left.\frac{\partial U'}{\partial \delta_{qi}}\right|_{\delta_{qi}=\overline{\delta_{qi}}}=0
	\end{align}
	and perform a Taylor expansion of $U'$ about it. The coefficient of the quadratic term $\delta_{qi}^2$ then becomes the effective Josephson energy $\tilde{E}_{Ji}, \ i\in\{1,2\}$ and the charging energy is defined as
	\begin{align}
	\left\{ 
	\begin{array}{cc}
	E_{C_{1}} =&\frac{e^2}{2}\bm{C_0}^{-1}[3,3],  \\
	E_{C_{2}} =&\frac{e^2}{2}\bm{C_0}^{-1}[4,4],  \\
	\end{array}
	\right.
	\end{align}
	where $\bm{C_0}^{-1}[i,j]$ is the element in the $i$-th row and $j$-th column of $\bm{C_0}^{-1}$. Next we can introduce bosonic creation ($\hat{a}_i^\dagger$) and annihilation operators $\hat{a}_i$ as
	\begin{align}
	\left\{ 
	\begin{array}{cc}
	\varphi_i  =& \sqrt{\frac{\hbar Z_{i}}{2}}\left(\hat{a}_i+{\hat{a}_i^\dagger}\right), \\
	Q_i  =& -i\sqrt{\frac{\hbar}{2Z_{i}}}\left(\hat{a}_i-{\hat{a}_i^\dagger}\right),
	\end{array}
	\right.
	\label{eq:quantization}
	\end{align}
	where
	\begin{align}
	\left\{ 
	\begin{array}{cc}
	Z_{r1} =& \frac{4e^2}{\hbar}\sqrt{L_1\bm{C_0}^{-1}[1,1]}, \\
	Z_{r2} =& \frac{4e^2}{\hbar}\sqrt{L_2\bm{C_0}^{-1}[2,2]}, \\
	Z_{q1} =& \sqrt{\frac{8E_{C1}}{\tilde{E}_{Jq1}}}, \\
	Z_{q2} =& \sqrt{\frac{8E_{C2}}{\tilde{E}_{Jq2}}}. \\
	\end{array}
	\right.
	\end{align}
	The resulting quantized circuit Hamiltonian becomes (under Rotating Wave Approximation)
	\begin{multline}
	H = \omega_{q1}\hat{a}_{q1}^{\dagger}\hat{a}_{q1} + \omega_{q2}\hat{a}_{q2}^{\dagger}\hat{a}_{q2} + \omega_{r1}\hat{a}_{r1}^{\dagger}\hat{a}_{r1}+\omega_{r2}\hat{a}_{r2}^{\dagger}\hat{a}_{r2} + \frac{\alpha_{q1}}{2}\hat{a}_{q1}^{\dagger}\hat{a}_{q1}\left(\hat{a}_{q1}^{\dagger}\hat{a}_{q1}-1\right)+ \frac{\alpha_{q2}}{2}\hat{a}_{q2}^{\dagger}\hat{a}_{q2}\left(\hat{a}_{q2}^{\dagger}\hat{a}_{q2}-1\right)\\
	g_{q1r1}\left(\hat{a}_{q1}^{\dagger}\hat{a}_{r1}+\hat{a}_{r1}^{\dagger}\hat{a}_{q1}\right) + g_{q2r2}\left(\hat{a}_{q2}^{\dagger}\hat{a}_{r2}+\hat{a}_{r2}^{\dagger}\hat{a}_{q2}\right) + g_{q1q2}\left(\hat{a}_{q1}^{\dagger}\hat{a}_{q2}+\hat{a}_{q2}^{\dagger}\hat{a}_{q_1}\right),
	\label{eq:total_H}
	\end{multline}
	where the coupling strengths are given by
	\begin{align}
	\left\{ 
	\begin{array}{cc}
	g_{q1r1} =& \frac{1}{2\sqrt{Z_1Z_{q_1}}}\bm{C_0}^{-1}[1,3],\\
	g_{q2r2} =& \frac{1}{2\sqrt{Z_2Z_{q_2}}}\bm{C_0}^{-1}[2,4],\\
	g_{q1q2} =& \frac{1}{2\sqrt{Z_{q_1}Z_{q_2}}}\bm{C_0}^{-1}[3,4]-\frac{L_g}{2L_x^2}\sqrt{Z_{q_1}Z_{q_2}}.
	\end{array}
	\right.
	\label{eq:coupling_strength1}
	\end{align}

	\section{Device parameters}
	
	Table.~\ref{table:optical_parameter} shows the capacitances and Josephson energies of different components of the device. Here, $E_{Jq1}$ and $E_{Jq2}$ represent the Josephson energies of the two qubits while $E_{Jc1}$ and $E_{Jc2}$ represent the Josephson energies of the two coupler junctions. The Josephson energies are computed from the Ambegaokar-Baratoff formula~\cite{Ambegaokar1963JJ_relation} $E_J=\frac{\pi\hbar\Delta}{4e^2R}$, where $\hbar$ is the reduced Planck constant and $e$ is the electronic charge. The low temperature resistances $R$ is calculated using the measured room temperature resistances $R'$ of identical test junctions with the assumption that $R=1.15R'$. For Al-AlO$_x$-Al junctions, the superconducting energy gap $\Delta=1.764k_b T_c$, where $k_b$ is the Boltzmann constant, and $T_c=1.2$ K is critical temperature for Aluminum. The capacitances are obtained from ANSYS Q3D simulation with labels indicated in Fig.~\ref{fig:device}(b).
	
	\setlength{\tabcolsep}{1.2 mm}
	\begin{table*}[h]
		\centering
		
		\begin{tabular}{|c|c|c|c|c|c||c|c|}
			\hline
			\multicolumn{6}{|c||}{Capacitance (fF)} 
			& \multicolumn{2}{c|}{$E_J$ (GHz)} \\  \hline
			\textrm{$C_{r1}$} & 131.7 & \textrm{$C_{cq1}$} & 2.53 & \textrm{$C_{q12}$ (Geometric)} & 0.73 & \textrm{$E_{Jq1}$} & 10.70 \\ \hline
			\textrm{$C_{r2}$} & 120.3 & \textrm{$C_{cq2}$} & 2.13 & \textrm{$C_{q12}$} (Coupler) & 3.09 & \textrm{$E_{Jq2}$}  & 13.26 \\  \hline
			\textrm{$C_{q1}$} & 109.6 & \textrm{$C_{r1q2}$} & 0.08 & \multirow{2}{*}{\textrm{$C_{q12}$ (total)}} & \multirow{2}{*}{3.82} & \textrm{$E_{Jc1}$} & 7.90 \\ \cline{1-4}\cline{7-8}
			\textrm{$C_{q2}$} & 89.7 & \textrm{$C_{r2q1}$} & 0.07 & & & \textrm{$E_{Jc2}$} & 7.74\\ \hline
		\end{tabular}
		\caption{Estimated capacitance and Josephson energies of our device. Capacitance are obtained from ANSYS Q3D simulations, and Josephson energies are calculated from room-temperature resistance measurements of the test junctions.}
		\label{table:optical_parameter}
	\end{table*}

	\section{Tuning cross-Kerr interaction strength}
	
\begin{figure}[ht]
		\includegraphics[width=0.5\columnwidth]{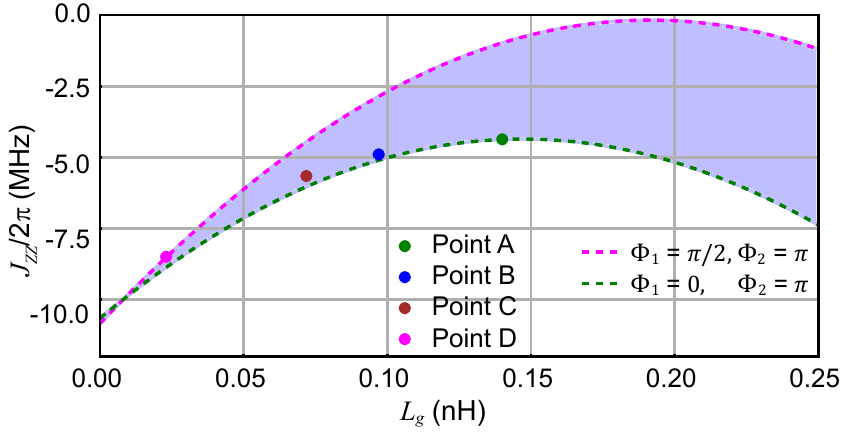}
		\caption{Cross-Kerr coupling strength as a function of shunt inductance $L_g$ to the ground for different flux biasing. The dashed magenta ($\Phi_1=\frac{\pi}{2}, \Phi_2=\pi$) and green ($\Phi_1=0, \Phi_2=\pi$) lines represent the boundaries within which we can operate. The experiments are performed at points A, B, and C.}
		\label{fig:Lg}
		\centering    
	\end{figure}

	According to Eq.~\ref{eq:coupling_strength1}, the exchange interaction strength $g_{q1q2}$ between the two qubit modes can be tuned by changing the inductance $L_g$, which in turn controls the \textit{ZZ} coupling. $L_g$ is further controlled by the external flux $\Phi_{ext}$ threading the small SQUID loop on the ground shunt. In this experiment, $\Phi_{ext}$ is a function of $\Phi_{1}, \Phi_{2}$. Fixing $\Phi_{2}=\pi$ and sweeping $\Phi_{1}$ one can tune $L_g$ and therefore modify the \textit{ZZ} coupling. In Fig.~\ref{fig:Lg} we show the range of \textit{ZZ} coupling strength ($J_{zz}/2\pi$) obtained by numerically diagonalizing Eq.~\eqref{eq:total_H} when $L_g$ is independently tuned between the operating conditions $\Phi_{1} = 0$ (green curve) and $\Phi_1=\pi/2$ (pink curve) while fixing $\Phi_2=\pi$. The experimentally determined values of cross-Kerr at points A ($J_{zz}/2\pi=-4.37$ MHz) and D ($J_{zz}/2\pi=-8.37$ MHz) are used to determine the boundaries of the shaded area. We have performed experiments at the biasing points A, B and C which correspond to $J_{zz}/2\pi=-4.37$, $-4.90$ and $-5.66$ MHz. Table.~\ref{table:device1} shows relevant Josephson energies and coupling strengths at the operating point A. Josephson energies are extracted by fitting experimentally obtained qubit frequencies, anharmonicities, cavity frequencies and cross-Kerr coupling using the Hamiltonian in Eq.~\ref{eq:total_H}.

	\setlength{\tabcolsep}{1.2 mm}
	\begin{table*}[h]
		\centering
		\begin{tabular}{|c|c|c|c|}
			\hline
			\multicolumn{4}{|c|}{Point A device parameters(GHz)} \\ \hline
			\textrm{$\tilde{E}_{Jq1}$} & 10.67 & \textrm{$g_{q1r1}/2\pi$} & 0.051 \\ \hline
			\textrm{$\tilde{E}_{Jq2}$} & 13.14 & \textrm{$g_{q2r2}/2\pi$} & 0.059 \\  \hline
			\textrm{$E_{Cq1}$} & 0.152 & \textrm{$g_{q1q2}/2\pi$} & 0.046 \\ \hline
			\textrm{$E_{Cq2}$} & 0.185   \\  \cline{1-2}
		\end{tabular}
		\caption{Effective Josephson energies, charging energies and coupling strengths at operating point A estimated from fitting experimental data to the quantized Hamiltonian in Eq.~\eqref{eq:total_H}.}
		\label{table:device1}
	\end{table*}

	\section{Evolution matrices for a two-qubit system}
	\label{supp:evol_mat}
	We consider a two-qubit system with static \textit{ZZ} coupling
	\begin{equation}
	H_0 = \omega_1 \ket{e_1}\bra{e_1} + \omega_2 \ket{e_2}\bra{e_2} + J_{zz} \ket{e_2 e_1} \bra{e_2 e_1},
	\end{equation}
	which is driven with the drive Hamiltonian
	\begin{equation}
	H = A_1 \sin(\omega_1 t - \phi_1) \sigma_{x_1} + A_2 \sin(\omega_2 t - \phi_2) \sigma_{x_2}.
	\end{equation}
	Going to the interaction frame rotating at $H_0$ and applying rotating wave approximation the full Hamiltonian takes the following matrix form
	\begin{equation}
	H_{\rm RWA} =
	\begin{bmatrix}
	0 & -i \frac{A_2}{2} e^{-i\phi_2} & -i \frac{A_1}{2} e^{-i\phi_1} & 0 \\
	i \frac{A_2}{2} e^{i\phi_2} & 0 & 0 & -i \frac{A_1}{2} e^{-i (J_{zz}(t + t_0) + \phi_1)} \\
	i \frac{A_1}{2} e^{i\phi_1} & 0 & 0 & -i \frac{A_2}{2} e^{-i (J_{zz}(t + t_0) + \phi_2)} \\
	0 & i \frac{A_1}{2} e^{i (J_{zz}(t + t_0) + \phi_1)} & i \frac{A_2}{2} e^{i (J_{zz}(t + t_0) + \phi_2)} & 0
	\end{bmatrix},
	\end{equation}
	where the drives are applied at time $t=t_0$. One can solve the Schrodinger equation $i\hbar \frac{\partial \ket{\psi}}{\partial t} = H_{\rm RWA} \ket{\psi}$ analytically when a drive to one qubit is applied at a time. The evolution of a generic two-qubit state can then be expressed as $\ket{\psi(t+t_0)} = F_i(t) \ket{\psi(t_0)}$ with 
	\begin{multline}
	\label{eq:F1}
	F_1(t,t_0,\phi_1) = \\
	\begin{bmatrix}
	\cos\left(\frac{A_1 t}{2}\right) & 0 &-e^{-i\phi_1}\sin(\frac{A_1t}{2}) & 0 \\
	0 & e^{\frac{i}{2}J_{zz}t}\left(\cos\left(\frac{\Omega_1 t}{2}\right) - i\frac{J_{zz}}{\Omega_1}\sin\left(\frac{\Omega_1 t}{2}\right)\right) & 0 &-\frac{A_1}{\Omega_1}{e^{\frac{i}{2}(J_{zz}(t+2t_0)-2\phi_1)}} \sin\left(\frac{\Omega_1 t}{2}\right) \\
	e^{i\phi_1}\sin(\frac{A_1t}{2}) & 0 &\cos\left(\frac{A_1t}{2}\right) & 0 \\
	0 &\frac{A_1}{\Omega_1}{e^{-\frac{i}{2}(J_{zz}(t+2t_0)-2\phi_1)}}\sin\left(\frac{\Omega_1 t}{2}\right) & 0 & e^{-\frac{i}{2}J_{zz}t}\left(\cos\left(\frac{\Omega_1 t}{2}\right)+i\frac{J_{zz}}{\Omega_1}\sin\left(\frac{\Omega_1 t}{2}\right)\right)
	\end{bmatrix}
	\end{multline}
	being the evolution matrix for qubit 1 and similarly,
	\begin{multline}
	\label{eq:F2}
	F_2(t,t_0,\phi_2) = \\
	\begin{bmatrix}
	\cos\left(\frac{A_2t}{2}\right) & -e^{-i\phi_2}\sin(\frac{A_2t}{2}) & 0 & 0 \\
	e^{i\phi_2}\sin(\frac{A_2t}{2}) & \cos\left(\frac{A_2t}{2}\right) & 0 & 0 \\
	0 & 0 & e^{\frac{i}{2}J_{zz}t}\left(\cos\left(\frac{\Omega_2 t}{2}\right) - i\frac{J_{zz}}{\Omega_2}\sin\left(\frac{\Omega_2 t}{2}\right)\right) & -\frac{A_2}{\Omega_2}{e^{\frac{i}{2}(J_{zz}(t+2t_0)-2\phi_2)}}\sin\left(\frac{\Omega_2 t}{2}\right) \\
	0 & 0 &\frac{A_2}{\Omega_2} {e^{-\frac{i}{2}(J_{zz}(t+2t_0) - 2\phi_2)}} \sin\left(\frac{\Omega_2 t}{2}\right) & e^{-\frac{i}{2}J_{zz}t} \left(\cos\left(\frac{\Omega_2 t}{2}\right)+i\frac{J_{zz}}{\Omega_2}\sin\left(\frac{\Omega_2 t}{2}\right)\right)
	\end{bmatrix}
	\end{multline}
	being the evolution matrix for qubit 2. Here, $\Omega_i = \sqrt{A_i^2 + J_{zz}^2}$, $t$ is the duration of the constant drive and $\phi_i$ determines the axis of rotation. Typically, pulses of duration $t=\frac{\pi}{2A_i}$ are applied (which correspond to $\pi/2$ rotation when the partner qubit is in $\ket{g}$) about $x$- ($\phi_i=-\pi/2$) and $y$-axes ($\phi_i=0$) during tomography. For example, starting from an initial state $\ket{\psi(t_0)}$, a pulse of duration $\frac{\pi}{2A_1}$ on qubit 1 about $x$-axis followed by a pulse of duration $\frac{\pi}{2A_2}$ on qubit 2 about $y$-axis will lead to the final state

	\begin{equation}
	\ket{\psi\left(\pi/2(A_2^{-1} + A_1^{-1}) + t_0\right)} = F_2\left(\frac{\pi}{2A_2}, t_0+\frac{\pi}{2A_1},0\right) F_1\left(\frac{\pi}{2A_1}, t_0,-\pi/2\right) \ket{\psi(t_0)}.
	\end{equation}
	For the case of rectangular pulses, nine evolution matrices are computed using Eqs.~\eqref{eq:F1} and \eqref{eq:F2} and used to obtain the coincidence counts.

	\section{Other states}
	\label{supp:entangled_state}
	
	The sate $\ket{\Psi_{\rm A}}$ in Fig.~2(a) of the main text is prepared by applying a $(\pi/2)_y$ rotation on qubit 1 followed by the same on qubit 2 starting from $\ket{gg}$. However, instead of a product state $(\ket{g}+\ket{e})(\ket{g}+\ket{e})/2$, due to finite \textit{ZZ} coupling, $\ket{\Psi_{\rm A}}$ becomes an entangled state as shown in table~\ref{table:entangled_states} (top panel). We prepare another set of entangled states $\ket{\Psi_{\rm B}}$ by waiting for a period of $\pi/J_{zz}$ after preparing $\ket{\Psi_{\rm A}}$. This waiting period works as controlled phase gate flipping the coefficient of the $\ket{ee}$ component as shown in the bottom panel of Table.~\ref{table:entangled_states}.

	\begin{table}[h]
		\centering
		\begin{tabular}{||c|c|c||}
			
			\hline
			$J_{zz}/2\pi$ (MHz) & $\ket{\Psi_{\rm A}}$ & Concurrence \\
			\hline
			$-4.37$ & $0.5\ket{gg} + 0.5\ket{ge} + (0.530+0.085i)\ket{eg} + (0.355-0.292i)\ket{ee})$ & 0.415 \\
			\hline
			$-4.90$ & $0.5\ket{gg} + 0.5\ket{ge} + (0.537+0.093i)\ket{eg} + (0.323-0.313i)\ket{ee})$ & 0.459 \\
			\hline
			$-5.66$ & $0.5\ket{gg} + 0.5\ket{ge} + (0.549+0.104i)\ket{eg} + (0.273-0.337i)\ket{ee})$ & 0.519 \\
			\hline
		\end{tabular}
		\\[2mm]
		\begin{tabular}{||c|c|c||}
			
			\hline
			$J_{zz}/2\pi$ (MHz) & $\ket{\Psi_{\rm B}}$ & Concurrence \\
			\hline
			$-4.37$ & $0.5\ket{gg} + 0.5\ket{ge} + (0.530+0.085i)\ket{eg} - (0.355-0.292i)\ket{ee}$ & 0.910 \\
			\hline
			$-4.90$ & $0.5\ket{gg} + 0.5\ket{ge} + (0.537+0.093i)\ket{eg} - (0.323-0.313i)\ket{ee})$ & 0.888 \\
			\hline
			$-5.66$ & $ 0.5\ket{gg} + 0.5\ket{ge} + (0.549+0.104i)\ket{eg} - (0.273-0.337i)\ket{ee})$ & 0.855 \\
			\hline
		\end{tabular}
		\caption{Explicit expressions for states $\ket{\Psi_{\rm A}}$ and $\ket{\Psi_{\rm B}}$ used in Fig.~2(a) of the main text and corresponding concurrences. 50 ns long rectangular $\frac{\pi}{2}$ pulses are used for these calculations.}
		\label{table:entangled_states}
	\end{table}

	\section{State fidelity comparison for different cross-Kerr strengths}
	\label{supp:state_fid}
	
\begin{figure*}[h]
		\centering
		\includegraphics[width=\textwidth]{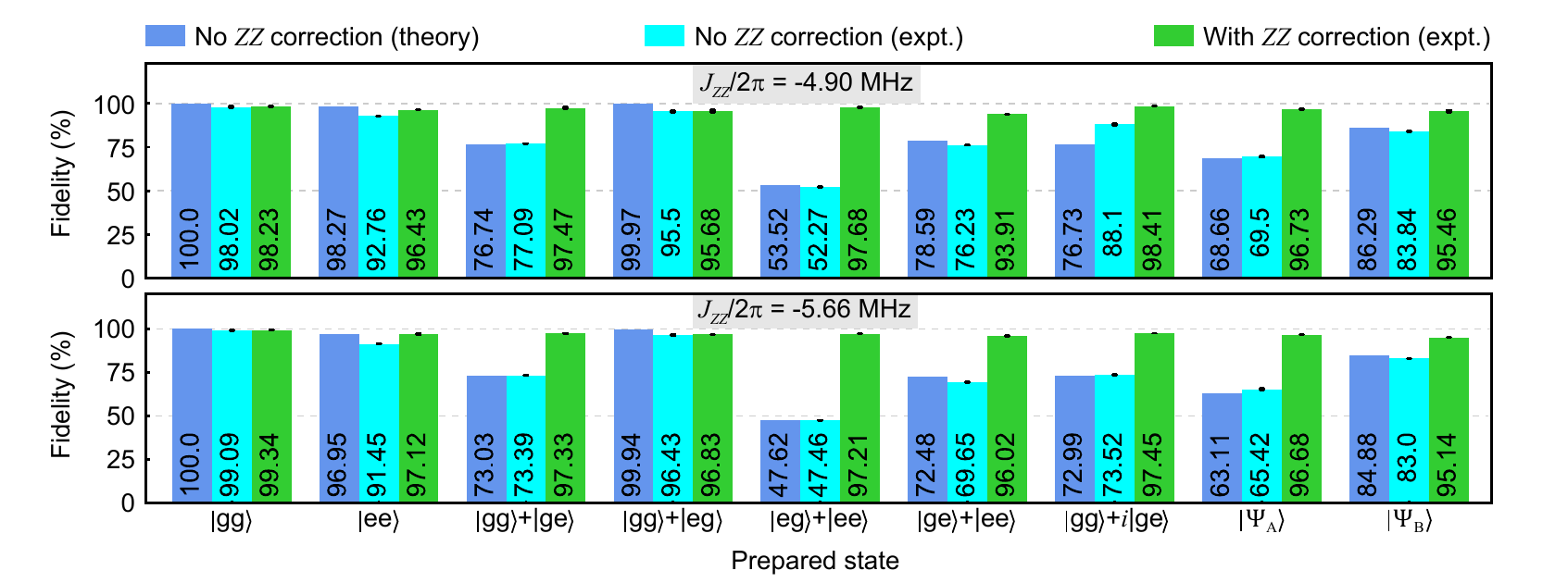}
		\caption{State fidelity comparison for different cross-Kerr strengths. Normalization coefficients are not shown for brevity. The expressions for the entangled states $\ket{\Psi_{\rm A}}$ and $\ket{\Psi_{\rm B}}$ are shown in Supplementary section~\ref{supp:entangled_state}.}
		\label{fig:state_fid}
	\end{figure*}

	The CCT works for arbitrary \textit{ZZ} coupling strengths. Comparison of state fidelities between regular tomography and CCT for $J_{zz}=-4.90$ and $-5.66$ MHz are shown in Fig.~\ref{fig:state_fid}. For all cases, CCT can reconstruct the states with larger than 95\% fidelity.
	
	\section{Error analysis}
	\label{supp:error}
	
	\begin{table}[ht]
		\scalebox{1.0}{%
			\centering
			\begin{tabular}{||c|c|c|c|c|c||}
				
				\hline
				\multicolumn{6}{||c||}{Error estimation for rectangular pulses} \\ 
				\hhline{||=|=|=|=|=|=||}
				$L_{\pi/2}=10$ ns & $\frac{\pi}{2}$ at $\ket{g}$ calibration & $\frac{\pi}{2}$ at $\ket{e}$ calibration & $\ket{f}$ leakage & $T_1$ decay & Average \\
				\hline
				$q1$ & $0.46\%$ & $1.60\%$ & $5.76\%$ & $0.03\%$ & $5.99\%$ \\
				\hline 
				$q2$ & $0.27\%$ & $1.77\%$ & $2.89\%$ & $0.08\%$ & $3.40\%$ \\
				\hline
				$\epsilon_{\rm tot}$ &\multicolumn{5}{|c||}{$9.19\%$} \\
				\hhline{||=|=|=|=|=|=||}
				$L_{\pi/2}=20$ ns & $\frac{\pi}{2}$ at $\ket{g}$ calibration & $\frac{\pi}{2}$ at $\ket{e}$ calibration & $\ket{f}$ leakage & $T_1$ decay & Average \\
				\hline
				$q1$ & $0.32\%$ & $1.60\%$ & $1.73\%$ & $0.06\%$ & $2.38\%$ \\
				\hline 
				$q2$ & $1.27\%$ & $1.61\%$ & $0.15\%$ & $0.16\%$ & $2.06\%$ \\
				\hline
				$\epsilon_{\rm tot}$ &\multicolumn{5}{|c||}{$4.39\%$} \\
				\hhline{||=|=|=|=|=|=||}
				$L_{\pi/2}=100$ ns & $\frac{\pi}{2}$ at $\ket{g}$ calibration & $\frac{\pi}{2}$ at $\ket{e}$ calibration & $\ket{f}$ leakage & $T_1$ decay & Average \\
				\hline
				$q1$ & $0.09\%$ & $3.69\%$ & $0.36\%$ & $0.32\%$ & $3.72\%$ \\
				\hline 
				$q2$ & $0.53\%$ & $1.61\%$ & $0.58\%$ & $0.82\%$ & $1.97\%$ \\
				\hline
				$\epsilon_{\rm tot}$ &\multicolumn{5}{|c||}{$5.62\%$} \\
				\hline 
		\end{tabular}}
		\\[2mm]
		\scalebox{1.0}{%
			\centering
			\begin{tabular}{||c|c|c|c|c|c||}
				\hline
				\multicolumn{6}{||c||}{Error estimation for Gaussian pulses} \\
				\hhline{||=|=|=|=|=|=||}
				$L_{\pi/2}=20$ ns & $\frac{\pi}{2}$ at $\ket{g}$ calibration & $\frac{\pi}{2}$ at $\ket{e}$ calibration & $\ket{f}$ leakage & $T_1$ decay & Average \\
				\hline
				$q1$ & $0.89\%$ & $1.42\%$ & $2.93\%$ & $0.06\%$ & $3.38\%$ \\
				\hline 
				$q2$ & $1.49\%$ & $0.74\%$ & $0.78\%$ & $0.16\%$ & $1.84\%$ \\
				\hline
				$\epsilon_{\rm tot}$ &\multicolumn{5}{|c||}{$5.16\%$} \\
				\hhline{||=|=|=|=|=|=||}
				$L_{\pi/2}=200$ ns & $\frac{\pi}{2}$ at $\ket{g}$ calibration & $\frac{\pi}{2}$ at $\ket{e}$ calibration & $\ket{f}$ leakage & $T_1$ decay & Average \\
				\hline
				$q1$ & $0.68\%$ & $2.47\%$ & $1.43\%$ & $0.63\%$  & $3.01\%$ \\
				\hline 
				$q2$ & $0.78\%$ & $0.80\%$ & $0.67\%$ & $1.63\%$ & $2.08\%$ \\
				\hline
				$\epsilon_{\rm tot}$ &\multicolumn{5}{|c||}{$5.03\%$} \\
				\hline
		\end{tabular}}
		\caption{Tomography error analysis for rectangular (top panel) and Gaussian (bottom panel) pre-rotation pulses. The dominant source of error for fast tomography pulses is the $\ket{f}$ leakage, explaining the fidelity drop in Fig.~2(b) of the main text. The maximum estimated error for slower pulses saturates to around 5\%.}
		\label{table:error}
	\end{table}
	
	We consider three error sources in tomography --- (1) relaxation ($T_1$) error (2) leakage to the $\ket{f}$ state during the application of tomography pulses and (3) imperfect calibration of the ${\pi}/{2}$ pulses that limit the performance of our tomography. Leakage to the state $\ket{f}$ is approximated through measuring the $\ket{e}\leftrightarrow \ket{f}$ Rabi oscillation amplitude after two $\pi$ pulses at $\ket{g}\leftrightarrow \ket{e}$ transition. The calibration error of the $\pi/2$ pulse depends on the other qubit's state, and two orthogonal states $\ket{g}$ and $\ket{e}$ are selected as two independent ${\pi}/{2}$ calibration error sources. We repeat the $\pi/2$ rotation 64 times and measure the residual population at $\ket{e}$ to calculate the calibration error. This method separates the $\ket{f}$ leakage contributions to the residual excitation (which happens at energy levels above $\ket{e}$). Each qubit's error $\epsilon_{qi}$ is the root mean square of all independent error sources, and the total error is $\epsilon_{\rm tot} = 1-(1-\epsilon_{q1})\cdot(1-\epsilon_{q2})$. Table.~\ref{table:error} shows the error analysis for rectangular and Gaussian pulses for different pulse lengths $L_{\pi/2}$. Note that in case of rectangular pulses, significant error happens for fast pulses ($L_{\pi/2}<10$ ns). Gaussian pulses have cutoffs at $\pm2\sigma$ so that $L_{\pi/2}=4\sigma$. In the whole range of Gaussian pulses, there is no significant difference between fast and slow pulses.

	\section{Extension to multi-qubit systems}
	\label{supp:three_qubits} 
	
	\begin{figure}[t]
		\centering\includegraphics[width=0.5\columnwidth]{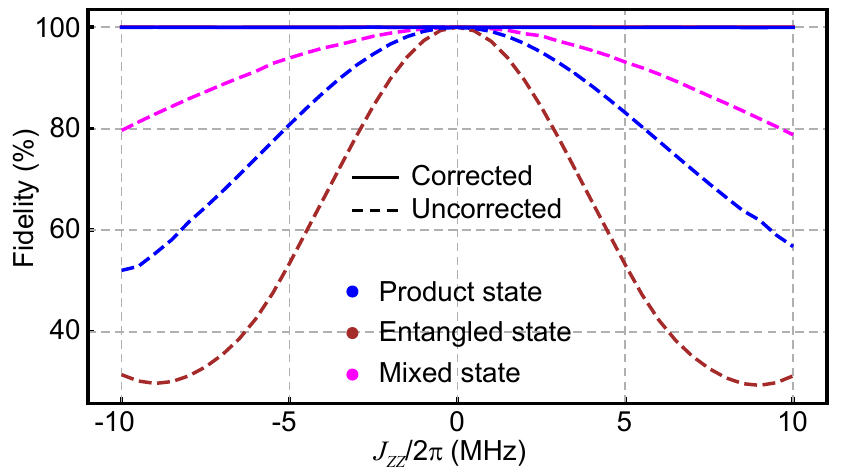}
		\caption{Comparison of fidelities from simulated three-qubit tomography when \textit{ZZ} coupling strength is varied for three diferent states. Dashed lines represent fidelities obtained with standard MLE and solid lines show fidelities with \textit{ZZ} correction using CCT.}
		\label{fig:3q}
	\end{figure}
	
	Our technique can theoretically be applied to a system having arbitrary number of qubits with different types of coupling, e.g. \textit{ZX} or \textit{XX} or any combination. As an example, we perform simulations for a three-qubit system with pairwise \textit{ZZ} coupling. The system is described by the Hamiltonian
	\begin{equation}
	H_{3q} = \sum_{i=1}^{3}\omega_{qi}\ket{e_i}\bra{e_i} + \sum_{1\leq i<j\leq 3}J_{ij}\ket{e_j e_i}\bra{e_j e_i}.
	\label{eq:3q}
	\end{equation}
	Fig.~\ref{fig:3q} shows the fidelity improvement for three test states --- a product states $(\ket{g}+\ket{e})^{\otimes 3}/ 2\sqrt{2}$, an entangled state $(\ket{ggg}+\ket{eee})/\sqrt{2}$, and a randomly chosen mixed state $\rho = 0.8\ket{\psi_{3a}}{\bra{\psi_{3a}}} + 0.2\ket{\psi_{3b}}{\bra{\psi_{3b}}}$ with $\ket{\psi_{3a}} = (\ket{ggg} - \ket{gge} + i\ket{geg} + \ket{gee} + \ket{egg})/\sqrt{5}$ and $\ket{\psi_{3b}} = (\ket{egg} + i\ket{ege} - \ket{eeg} + \ket{eee})/2$ when the \textit{ZZ} correction is applied through CCT. For simplicity, we have set all three stray \textit{ZZ} coupling between qubits the same $J_{zz} = J_{12} = J_{23} = J_{13}$, however, CCT is independent of this choice.

	\section{Measurement error mitigation}
	\label{supp:meas_error}
\begin{figure*}[h]
		\centering\includegraphics[width=0.5\columnwidth]{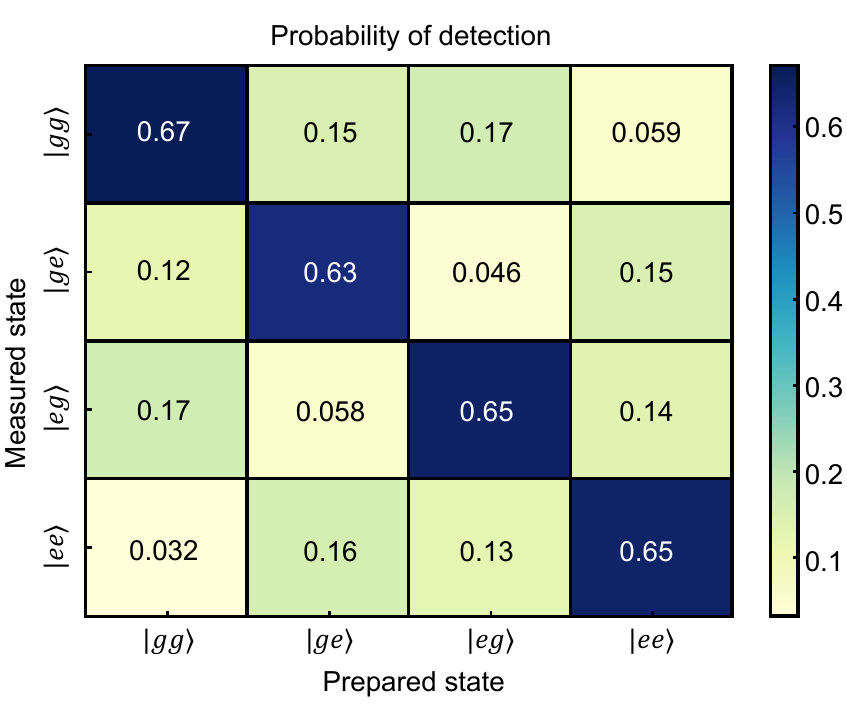}
		\caption{The heatmap of a typical measurement confusion matrix.}
		\label{fig:heatmap}
	\end{figure*}

	We perform simultaneous readout of both qubits and thus each measurement reveals two bits of information corresponding to projection in $\ket{gg}, \ket{ge}, \ket{eg}$ or $\ket{ee}$ state. We assume that our state preparation is perfect and the error is associated with the measurement. In order to characterize the measurement error, we prepare each of the four basis states 5000 times followed by immediate measurements to construct the confusion matrix $\overleftrightarrow{M}$. Each element $M_{j,k}$ of the confusion matrix denotes the probability of obtaining a basis state $\ket{j}$ when a basis state $\ket{k}$ is prepared. Now, for a given state, the experimentally obtained probability distribution $\vec{P}_{\rm expt}$ will be skewed due to measurement error as $\vec{P}_{\rm expt} = \overleftrightarrow{M} \vec{P}_{\rm id}$, where $\vec{P}_{\rm id}$ is the probability distribution in the absence of measurement error. In order to mitigate this error we invert the confusion matrix and obtain the ideal distribution $\vec{P}_{\rm id} = \overleftrightarrow{M}^{-1} \vec{P}_{\rm expt}$ which is used for the tomography. The heat-map of a typical confusion matrix obtained with a repetition of 5000 measurements for each prepared state is shown in Fig.~\ref{fig:heatmap}.
	
	\input{supp_ZZ_arxiv.bbl}

\end{document}

%% file: main_ZZ_arxiv.bbl
%

%% file: supp_ZZ_arxiv.bbl
%